# Spin-orbit torques in heavy metal/ferromagnet bilayers with varying strength of interfacial spin-orbit coupling

Lijun Zhu,[1]* D. C. Ralph,[1,2] and R. A. Buhrman[1]
1. Cornell University, Ithaca, NY 14850
2. Kavli Institute at Cornell, Ithaca, New York 14853, USA
*e-mail: lz442@cornell.edu

Despite intense efforts it has remained unresolved whether and how interfacial spin-orbit coupling (ISOC) affects spin transport across heavy metal (HM)/ferromagnet (FM) interfaces. Here we report conclusive experiment evidence that the ISOC at HM/FM interfaces is the dominant mechanism for "spin memory loss". An increase in ISOC significantly reduces, in a linear manner, the dampinglike spin-orbit torque (SOT) exerted on the FM layer via degradation of the spin transparency of the interface for spin currents generated in the HM. In addition, the fieldlike SOT is also dominated by the spin Hall contribution of the HM and decreases with increasing ISOC. This work reveals that ISOC at HM/FM interfaces should be *minimized* to advance efficient SOT devices through atomic layer passivation of the HM/FM interface or other means.

**Key words**: spin orbit torque, spin Hall effect, spin orbit coupling, spin transparency

Current-induced spin-orbit torques (SOTs) in heavy metal/ferromagnet (HM/FM) systems have attracted remarkable attention due to their potential for the efficient manipulation of magnetization in metals and insulators at the nanoscale [1-5]. When the spin Hall effect (SHE) of the HMs is the dominant source of the SOTs as is the case for many HM/FM systems [4,6], achieving high spin transparency ($T_{int}$) at the HM/FM interfaces is of the same importance as achieving a large spin Hall ratio ($\theta_{SH}$) in the HMs for obtaining a large dampinglike SOT efficiency per *unit bias current density* $\xi_{DL}^j \equiv T_{int}\theta_{SH}$, the quantity of primary importance for efficiently driving magnetization switching [1,2,4,6], magnetization dynamics, and chiral spin texture displacement [3]. Recently, it has become recognized that so-called "spin memory loss" (SML) at many HM/FM interfaces can degrade $T_{int}$ [7-9] and therefore substantially reduce the spin current that is pumped into the HM by a resonantly excited FM, or alternatively reduce the $\xi_{DL}^j$ exerted on the FM layer by spin current arising from the SHE in the HM. While intermixing and atomic disorder were initially expected to be the dominant cause of SML, an extended first-principles analysis of spin pumping by Zwierzycki *et al.*[10] indicates that the spin mixing conductance $G^{\uparrow\downarrow}$ of a HM/FM interface is insensitive to interfacial diffusion and atomic disorder. Thus a different explanation for SML is required.

The initial spin-pumping theories and the supporting experiments omitted any effects of the interfacial spin-orbit coupling (ISOC) at the HM/FM interface and assumed the (pumped) spin current to be continuous across the HM/FM interface [7,11-13]. More recently, a number of theoretical efforts [14,15] have been made to assess the possible relevance of ISOC to $T_{int}$ and the SOT efficiencies. Some calculations have suggested that ISOC at the HM/FM interface could contribute to SML [16,17], acting to enhance the spin scattering at the interface even in absence of inter-diffusion or atomic disorder [18]. However, clear experimental evidence for whether and how ISOC is relevant in SOTs and $T_{int}$ has been lacking, as it is a major challenge to engineer and quantify a changing strength of ISOC within one material system while keeping the SOT efficiencies high enough to measure accurately.

In this letter, we demonstrate that ISOC at Pt/Co and Au$_{0.25}$Pt$_{0.75}$/Co interfaces can be tuned significantly via thermal engineering of the spin-orbit proximity effect. From this ability we show that ISOC is the dominant mechanism for SML and reduces the dampinglike SOT efficiency by degrading $T_{int}$ in a linear manner. We also find that the fieldlike torque is dominated by the SHE of the HM and is affected by ISOC only via degradation of $T_{int}$. The ISOC-generated SOTs is found to be negligible in the studied HM/Co systems.

We studied several magnetic bilayers (see Supplemental Material [19]), each of which underwent repeated cycles of measurements and annealing (see Table 1) to tune the strength of ISOC. These include sample series Pt 4/Co 0.85 (P1-P4) and Au$_{0.25}$Pt$_{0.75}$ 4/Co 0.85 (P5-P8) with perpendicular magnetic anisotropy (PMA), and Pt 4/Co 3.2 (I1-I4) with in-plane magnetic anisotropy (IMA) (the numbers are layer thicknesses in nm). In addition, we measured two IMA reference samples Pt 4/Hf 0.67/Co 1.4 (R1) and Pt 4/Co 1.4 (R2).

We used two quantities, the interfacial magnetic anisotropy energy density ($K_s$) and the anomalous Hall conductivity ($\sigma_{AH}$), as indicators of the strength of ISOC. It has been well established that $K_s$ in HM/Co bilayers and superlattices originates from spin-orbit coupling (SOC)-enhanced orbital magnetic moments localized at the first layer adjacent to the interface [20,21]. A HM with a strong SOC (e.g. 5$d$ Pt) can modify the perpendicular orbital moments in an adjacent 3$d$ Co layer via strong interfacial 3$d$-5$d$ hybridization, which enhances the latter's spin-orbit interaction and thereby $K_s$. $\sigma_{AH}$ in thick Co films is determined by the intrinsic contribution from the internal spin-orbit interaction related to the integral of the Berry curvature over occupied states [22]. However, in ultrathin ferromagnetic films where the electron mean-free-path is determined by interfacial scattering, this "bulk" contribution to $\sigma_{AH}$ decreases linearly towards zero with decreasing FM thickness, which allows the contribution from a strong

ISOC to dominate [23]. We find that $\sigma_{AH}$ in ultrathin Co films in contact with Pt (or Au$_{0.25}$Pt$_{0.75}$) provides a separate and independent measure of the relative strength of ISOC. In Figs. 1(a) and 1(b) we plot $K_s$ and $\sigma_{AH}$ as measured [19] in the different samples and for different annealing treatments. We note that the dominant contributions to the magnitudes of $K_s$ (1-3.5 erg/cm$^2$) and $\sigma_{AH}$ and their variation with annealing come from the Pt/Co and Au$_{0.25}$Pt$_{0.75}$/Co interfaces and their ISOC variation, rather than the MgO interfaces or bulk contributions. The upper bound for the $K_s$ contribution from the Co-MgO interface is only ~0.6 erg/cm$^2$, i.e. $K_s$(MgO) as indicated by dotted line in Fig. 1(a). This bound was determined from a Pt/Hf/Co/MgO sample (R1) and assumes the Hf/Co interface has a negligible contribution to $K_s$. Upon 300°C annealing, $K_s$(MgO) from R1 changed from 0.56±0.05 to 0.66±0.10 erg/cm$^2$, which is a minimal variation compared to that of the Pt/Co/MgO samples (e.g. nearly doubling from I1 to I2 for a similar 300 C anneal; also see [19]). Similarly, an upper bound for the Co bulk contribution to $\sigma_{AH}$ is estimated to be 0.30×10$^4$ $\Omega^{-1}$ m$^{-1}$ and 1.12×10$^4$ $\Omega^{-1}$ m$^{-1}$ for the 0.85 and 3.2 nm "bulk" Co layers respectively, as indicated by $\sigma_{AH}$(Co) in Fig. 1(b), by using the $\sigma_{AH}$(Co) = 0.49×10$^4$ $\Omega^{-1}$ m$^{-1}$ as determined from R1 and the proportional dependence of the intrinsic $\sigma_{AH}$ on FM thickness [23]. There was also only a minimal effect of annealing on $\sigma_{AH}$ of R1 ($\sigma_{AH}$ = 0.46×10$^4$ $\Omega^{-1}$ m$^{-1}$ after 300°C annealing).

As is clearly seen in Fig. 1, the changes in $K_s$ and $\sigma_{AH}$, track each other fairly closely for each sample set, consistently indicating an evolution of ISOC with the annealing steps. Specifically, for the two PMA samples the ISOC first increases and then gradually drops back down, while for the IMA Pt/Co samples it goes up monotonically. We tentatively attribute the different behavior of the thicker IMA Co sample set to differences in strain relaxation that occur for different Co thicknesses, associated with the large lattice mismatch relative to Pt. We attribute the evolution of ISOC with the annealing steps to thermal tuning of the spin-orbit proximity effect [14,24]. It has been shown that a HM with strong SOC can modify the SOC in the interfacial region of an adjacent layer due to the spin-orbit proximity effect [14,24]. A particularly striking example is the enhanced SHE and quantum anomalous Hall effect in graphene [24,25]. The annealing-induced changes of $K_s$ and $\sigma_{AH}$ that we observe are unrelated to atomic inter-mixing at the HM/Co interfaces. Chemical depth profiling indicates displacement of the Co is at most at the single-atomic-layer level, while x-ray reflectivity indicates that annealing sharpens the HM/Co interfaces rather than inducing intermixing [19,26]. Detailed transport and magnetic characterization of the samples safely further exclude the occurrence of any substantial atomic inter-mixing in these samples. $K_s$ and $\sigma_{AH}$ are high in the as-grown state, and keep increasing with annealing for the IMA Pt/Co series (I1-I4), whereas an intermixed layer that is deliberately introduced (e.g. an alloy layer of Co$_{50}$Pt$_{50}$ or Pt$_3$Co/Co$_{50}$Pt$_{50}$/Co$_3$Pt) is found to substantially degrade $K_s$ and $\sigma_{AH}$ [19,27]. We note that our results are also consistent with the earlier finding that high-temperature annealing can separate Co out from initially chemically disordered CoPt alloys and therefore enhance the PMA of Pt/Co (111) superlattices [28].

In Figs. 1(c) and 1(d) we show the dampinglike and fieldlike SOT efficiencies for the samples before and after annealing as determined by harmonic response measurements [19,29,30]. Since the resistivities of both Pt and Au$_{0.25}$Pt$_{0.75}$ can change during the annealing process [19] and alter the SOT efficiencies per unit bias current density, $\xi^j_{\text{DL(FL)}}$, we show first the SOT efficiencies per *unit applied electric field*, $\xi^E_{\text{DL(FL)}} = \xi^j_{\text{DL(FL)}}/\rho_{\text{HM}}$ ( $\rho_{\text{HM}}$ is the HM resistivity), in Fig. 1(c), and then for completeness $\xi^j_{\text{DL(FL)}}$ in Fig. 1(d). The dependence on annealing is similar for both quantities, but in the following we will focus on $\xi^E_{\text{DL(FL)}}$ as being more fundamental for considering changes in intrinsic SOTs.

Focusing first on the technologically more important and stronger dampinglike component $\xi^E_{\text{DL}}$ for the PMA series Pt/Co (P1-P4) and Au$_{0.25}$Pt$_{0.75}$/Co (P5-P9), upon the first annealing step $\xi^E_{\text{DL}}$ drops by ~50% and then gradually recovers back to some extent as result of the two subsequent annealing steps. For the IMA Pt/Co series (I1-I4), $\xi^E_{\text{DL}}$ monotonically *decreases* with annealing. The key observation is that for all three cases the variations of $\xi^E_{\text{DL}}$ upon annealing are strikingly well (negatively) correlated with the variation of the ISOC strength as seen in both Figs. 1(a) and 1(b). The fact that $\xi^E_{\text{DL}}$ tracks the ISOC quantified by $K_s$ and $\sigma_{AH}$ in a markedly close manner strongly indicates that SML is mainly due to interfacial spin-orbit scattering. Our findings are consistent with the theoretical calculations [16-18] that a strong ISOC at the HM/FM interface enhances interfacial spin scattering. As schematically shown in Fig. 2(a), ISOC leads to a sharp drop and discontinuity of the spin current density at HM/FM interfaces in a SOT experiment.

In order to quantitatively determine the scaling of $\xi^E_{\text{DL}}$ with ISOC, in Figs. 2(b) and 2(c) we plot $\xi^E_{\text{DL}}$ as a function of the two more direct indicators of ISOC strength, $K_s^{\text{ISOC}} = K_s - K_s$(MgO) and $\sigma_{AH}^{\text{ISOC}} = \sigma_{AH} - \sigma_{AH}$(Co), for the three sample series. Strikingly, $\xi^E_{\text{DL}}$ scales roughly linearly with $K_s^{\text{ISOC}}$ and $\sigma_{AH}^{\text{ISOC}}$ for each sample series. We also note that the intercepts are apparently indicating the values of $\xi^E_{\text{DL}}$ in absence of the ISOC. It is rather remarkable that the best linear fits as a function of $K_s^{\text{ISOC}}$ ($\sigma_{AH}^{\text{ISOC}}$) for both the PMA and IMA Pt/Co sample series indicate a giant $\xi^E_{\text{DL}}$ of $\approx 6 \times 10^5$ $\Omega^{-1}$ m$^{-1}$ for an ideal Pt/Co interface with zero ISOC. We point out that this large value $\xi^E_{\text{DL}} \approx 6 \times 10^5$ $\Omega^{-1}$ m$^{-1}$ is still a lower bound for the internal spin Hall conductivity ($\sigma_{SH}$) of Pt, because the spin backflow (SBF) is not taken into account. If we assume the same SBF correction of $T^{\text{SBF}}_{\text{int}} \approx 0.48$ for the Pt/Co interface as determined by a recent Pt thickness dependent study [31], we obtain $\sigma_{SH} \approx 1.3 \times 10^6$ $\Omega^{-1}$ m$^{-1}$ for Pt. For the Au$_{0.25}$Pt$_{0.75}$/Co interface, the best linear fits of $\xi^E_{\text{DL}}$ vs $K_s^{\text{ISOC}}$ ($\sigma_{AH}^{\text{ISOC}}$) yield $\xi^E_{\text{DL}} \approx 7.5 \times 10^5$ $\Omega^{-1}$ m$^{-1}$ in the zero-ISOC limit. Taking into account $T^{\text{SBF}}_{\text{int}} \approx 0.63$ for Au$_{0.25}$Pt$_{0.75}$/Co interface as reported recently [32], we then obtain an internal value of $\sigma_{SH} \approx 1.2 \times 10^6$ $\Omega^{-1}$ m$^{-1}$ for Au$_{0.25}$Pt$_{0.75}$, comparable with that of Pt. Here we emphasize that the strong $\xi^E_{\text{DL}}$ variations are unlikely to be attributed to any annealing-induced change of either $G^{\uparrow\downarrow}$ or the spin Hall conductivity of the HM layers. First, calculations have indicated that $G^{\uparrow\downarrow}$ is insensitive to interfacial disorder and ISOC [14-16], although recent work has suggested [16] that SBF at HM/FM interface could be modified by the interfacial spin-orbit scattering. Second, $\sigma_{SH}$ for Pt and $Au_{1-x}Pt_x$ is dominated by the intrinsic SHE determined by the topology of the band structure, which for simple fcc metals is only dependent on the long-range crystal structure and is hence robust against localized changes in structural disorder that might occur during annealing [33]. We can also report that the strong variations in $\xi_{DL}^E$ are unrelated to atomic intermixing at the HM/Co interface. We found that inserting a 0.6 nm intermixed Co-Pt layer (i.e. $Pt_3Co$ 0.2/$Co_{50}Pt_{50}$ 0.2/$Co_3Pt$ 0.2) between the Pt and Co had no significant effect on $\xi_{DL}^E$ [19].

Our results indicate that the development of highly efficient HM/FM spin-torque devices, e.g. magnetic memories, will generally be advanced by *minimizing* ISOC at the HM/FM interfaces. As an example, we show in Figs. 3(a) and 3(b) that, through atomic Hf layer passivation of the HM/FM interface, the ISOC at a Pt/Co interface can be significantly reduced (samples R1 and R2), which beneficially results in an significant improvement of $\xi_{DL}^E$ (and hence $T_{int}$) and a considerable reduction (×4.6) in the ISOC enhancement of magnetic damping via SML [7-9,16] or/and two-magnon scattering [34].

Turning to the fieldlike torque, for all the samples $\xi_{FL}^E$ is negative (opposite to the Oersted field generated by the bias current) and small in magnitude compared to $\xi_{DL}^E$ (Fig. 1(c)). According to the SBF theory, the fieldlike torque generated by a SHE spin current impinging onto a HM/FM interface is expected to be quite weak in comparison to the dampinglike torque [14], scaling as $ImG^{\uparrow\downarrow} \ll ReG^{\uparrow\downarrow}$. While the magnitude of $|\xi_{FL}^E|$ relative to $\xi_{DL}^E$ for the Pt/Co and $Au_{.025}Pt_{0.75}$/Co samples is somewhat higher than predicted by these model calculations it is still comparatively low, consistent with other Pt/Co systems [9]. We find that $|\xi_{FL}^E|$ shows the same trend as that of $\xi_{DL}^E$ with annealing, which we take as strong evidence that the SHE is the dominant mechanism for both torque components in our PMA and IMA samples, with ISOC determining the SML at the HM/Co interface.

Recent theoretical work [14,15,17] has advanced the understanding that there are various mechanisms by which a strong ISOC might generate rather than degrade SOTs in HM/FM bilayers. However, we find no indication of any significant ISOC-induced SOTs that scale positively with ISOC strength. Therefore, we conclude that the SOTs in the HM/Co systems studied here are predominantly due to the SHE in the HMs while the ISOC-generated torques are negligible.

In summary, we have demonstrated that thermal engineering of the spin-orbit proximity effect is a simple but effective approach to controllably vary the strength of ISOC at Pt/Co and $Au_{25}Pt_{75}$/Co interfaces. From this variation, we have obtained conclusive experimental evidence that ISOC at a HM/FM interface is the dominant mechanism for SML and increasing ISOC significantly reduces the dampinglike and fieldlike SOTs by enhancing interfacial spin scattering and degrading $T_{int}$ in a linear manner. For both PMA and IMA HM/FM systems studied here, the ISOC-generated torques are negligible. This work also indicates that the development of highly efficient HM/FM spin-torque devices, e.g. magnetic memories, will generally be advanced by *minimizing* ISOC effects, through atomic layer passivation of the HM/FM interface or other means. The approach we utilized here to tune the ISOC strength should be also beneficial for research and applications of Dzyaloshinskii-Moriya interaction (skyrmion and chiral domain wall physics) and other ISOC effects.

This work was supported in part by the Office of Naval Research and by the NSF MRSEC program (DMR-1719875) through the Cornell Center for Materials Research. Support was also provided by the Office of the Director of National Intelligence (ODNI), Intelligence Advanced Research Projects Activity (IARPA), via contract W911NF-14-C0089. The views and conclusions contained herein are those of the authors and should not be interpreted as necessarily representing the official policies or endorsements, either expressed or implied, of the ODNI, IARPA, or the U.S. Government. The U.S. Government is authorized to reproduce and distribute reprints for Governmental purposes notwithstanding any copyright annotation thereon. Additionally this work was supported by the NSF (ECCS-1542081) through use of the Cornell Nanofabrication Facility/National Nanofabrication Infrastructure Network.

Table 1. Sample configurations and annealing conditions. Bilayers P1-P8 have perpendicular magnetic anisotropy, while others have in-plane magnetic anisotropy. Layer thicknesses for Co, Pt, $Au_{0.25}Pt_{0.75}$ and Hf are in nm.

| # | Bilayers | Anneal condition |
|---|---|---|
| P1 | Pt 4/Co 0.85 | As-grown |
| P2 | Pt 4/Co 0.85 | 350 ºC,2 h |
| P3 | Pt 4/Co 0.85 | 350 ºC,4 h |
| P4 | Pt 4/Co 0.85 | 350 ºC,4 h +400 ºC,1 h |
| P5 | $Au_{0.25}Pt_{0.75}$ 4/Co 0.85 | As-grown |
| P6 | $Au_{0.25}Pt_{0.75}$ 4/Co 0.85 | 350 ºC,2 h |
| P7 | $Au_{0.25}Pt_{0.75}$ 4/Co 0.85 | 350 ºC,4 h |
| P8 | $Au_{0.25}Pt_{0.75}$ 4/Co 0.85 | 350 ºC,4 h +400 ºC,1 h |
| I1 | Pt 4/Co 3.2 | As-grown |
| I2 | Pt 4/Co 3.2 | 300 ºC, 0.5 h |
| I3 | Pt 4/Co 3.2 | 300 ºC, 0.5 h +350 ºC,4 h |
| I4 | Pt 4/Co 3.2 | 300 ºC,0.5 h +350 ºC,4 h+450 ºC,1 h |
| R1 | Pt 4/ Hf 0.67/ Co 1.4 | As-grown/ 300 ºC, 0.5 h |
| R2 | Pt 4/ Co 1.4 | As-grown |

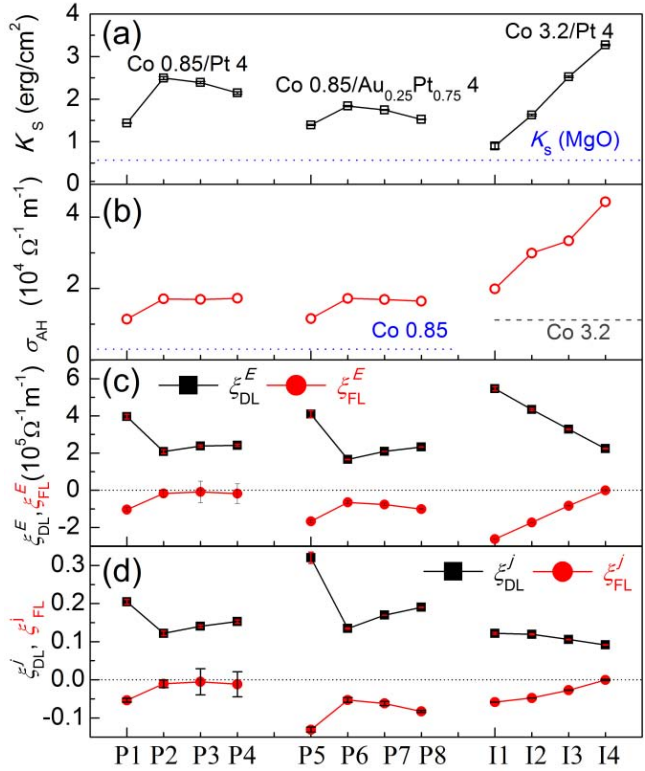

Fig. 1. (a) The interfacial magnetic anisotropy energy density ($K_s$), (b) The anomalous Hall conductivities ($\sigma_{AH}$), (c) The dampinglike and fieldlike SOT efficiencies per unit applied electric field, and (d) The dampinglike and fieldlike SOT efficiencies per unit bias current density for the sample series defined in Table 1. The blue dotted line in (a) represents an upper bound for $K_s$ from the Co/MgO interface as determined from measurement of the R1 reference sample with Hf passivation of the interface; the blue dotted and grey dashed lines in (b) represent calculated values of $\sigma_{AH}$ for a 0.85 nm Co layer and a 3.2 nm Co layer, respectively, in the absence of significant ISOC at the HM/Co interface.

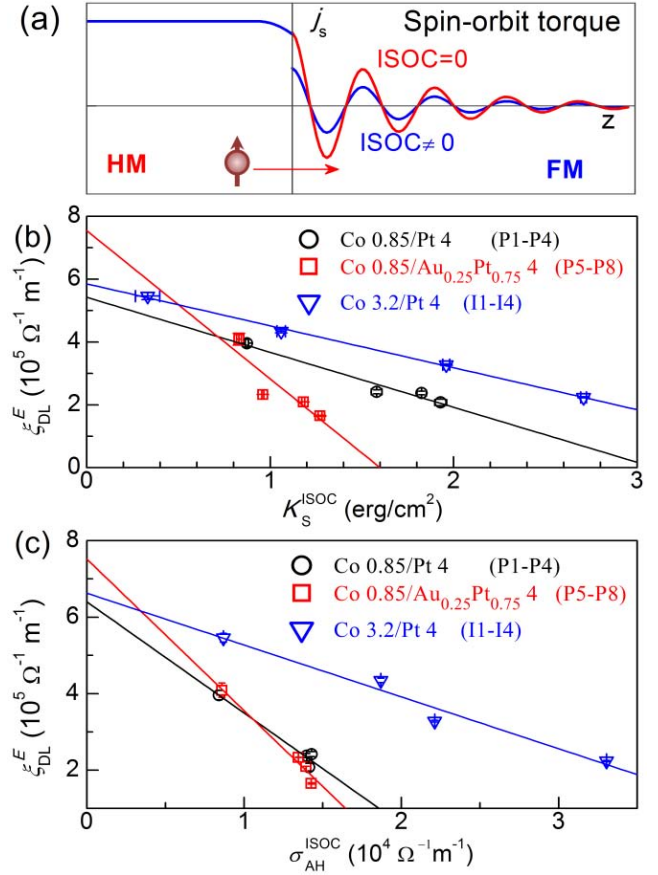

Fig. 2. (a) Schematic illustration of the z-profile of the spin current density ($j_s$) at a HM/FM interface in spin-orbit torque experiment, where the spin current arises from the SHE in the bulk of the HM. The blue and red curves represent $j_s$ for non-zero and zero ISOC at the HM/FM interface, respectively. The red arrow points to the spin transport direction. In the case of non-zero ISOC, $j_s$ drops when traveling across the interface. Scaling of dampinglike SOT efficiency per unit applied electric field with (b) the interfacial magnetic anisotropy energy density ($K_s^{ISOC}$) at the HM/Co interface, and (c) the HM/Co interface contribution to the anomalous Hall conductivity ($\sigma_{AH}^{ISOC}$), for the sample series defined in Table 1. The solid lines in (b) and (c) represent the best linear fits to the data.

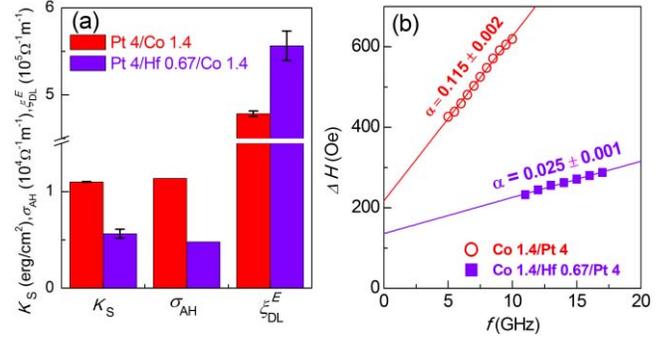

Fig. 3. (a) The interfacial magnetic anisotropy energy density ($K_s$), the anomalous Hall conductivities ($\sigma_{AH}$), and dampinglike SOT efficiency per unit applied electric field ($\xi_{FL}^E$), and (b) Frequency dependence of ferromagnetic resonance linewidth broadening ($\Delta H$) and magnetic damping constant ($\alpha$) for Pt 4/Co 4 and Pt 4/Hf 1.4/Co 4 as-grown samples.

Supplemental Material for

# Spin-orbit torques in heavy metal/ferromagnet bilayers with varying strength of interfacial spin-orbit coupling

L. J. Zhu, [1]* D.C. Ralph,[1,2] and R. A. Buhrman[1]
1. Cornell University, Ithaca, NY 14850
2. Kavli Institute at Cornell, Ithaca, New York 14853, USA

**1. Sample fabrication and characterizations**.

**2. Interfacial magnetic anisotropy and anomalous Hall conductivity**.

**3**. **Annealing effects on Pt/Co and Pt/Hf/Co interfaces**

**4. Spin-orbit torque efficiencies and resistivities**

**5. ISOC variations are unrelated to increased intermixing at the HM/Co interfaces**

**6 SOTs variations are unrelated to increased intermixing at the HM/Co interfaces**

**7. Direct structural characterizations of the HM/Co interfaces**

**8. Strain variation**

## 1. Sample fabrication and characterization.

The samples were deposited at room temperature by sputtering onto oxidized Si substrates with an argon pressure of 2 mTorr and a base pressure of ~$10^{-8}$ Torr. A 1 nm Ta underlayer was utilized to improve the adhesion and smoothness of the HM layer, i.e. Pt or $Au_{0.25}Pt_{0.75}$, but is expected to contribute negligible spin current into the Co layer as a consequence of the small thickness and high resistivity of Ta, the short spin diffusion length (< 2 nm) of Pt and $Au_{0.25}Pt_{0.75}$. Each sample was capped with a 2 nm MgO layer and a 1.5 nm Ta layer that was fully oxidized upon exposure to atmosphere. To minimize interfacial intermixing, each layer was sputtered at a low rate (e.g. 0.007 nm/s for Co and 0.016 nm/s for Pt) by introducing an oblique orientation of the target to the substrate and by using low magnetron sputter power. Annealing was performed in a vacuum furnace with a background pressure of ~$10^{-7}$ Torr. The samples were patterned into 5×60 μm² Hall bars and 10×20 μm² microstrips by ultraviolet photolithography and argon ion milling, followed by deposition of 5 nm Ti and 150 nm Pt as electrical contacts. A Quantum Design PPMS with a vibrating sample magnetometer (sensitivity~$10^{-7}$ emu) was used to determine the sample magnetization ($M_s$). A Signal Recovery DSP Lock-in Amplifier (Model 7265) was used to source a 4 V sinusoidal voltage onto the Hall bars and to detect the first and second harmonic Hall voltage responses, $V_{1\omega}$ and $V_{2\omega}$. All the measurements were performed at room temperature. The magnetic damping of the in-plane magnetized samples (10×20 μm² microstrips) was measured by spin-torque ferromagnetic resonance as a function of frequency.

## 2. Interfacial magnetic anisotropy and anomalous Hall conductivity

For a PMA HM/FM system, the effective PMA field ($H_k$) can be determined by fitting the in-plane bias field ($H$) dependence of the in-phase first harmonic Hall voltage response ($V_{1\omega}$) to the sinusoidal voltage ($V_{in}$ = 4 V) applied to the Hall bar following
$$V_{1\omega} = \pm V_{AH}\cos\theta = \pm V_{AH}(1-H^2/2H_k^2)$$
where $V_{AH}$ is the anomalous Hall voltage determined by measuring $V_{1\omega}$ as a function of $H_z$, $\theta \approx H/H_k \ll 1$ is the polar angle of the magnetic moment tilted away from the film normal direction, and the signs ± correspond to the initial magnetization states of $\pm M_z$, respectively. Here the charge current flows in the x-direction, $V_{1\omega}$ is measured along the transverse direction y, and z is the out-of-plane direction. We measured both $H_x$ and $H_y$ dependencies of $V_{1\omega}$ for the $\pm M_z$ cases (see Fig. S1(a)) and averaged the four values of $H_k$ for the bilayers P1-P8.

For IMA Pt 4/Co 3.2 (I1-I4), $dV_{1\omega}/dH_z$ is first determined under different in-plane bias fields $H_x$. $H_k$ and $V_{AH}$ were then determined by fitting the data to $dV_{1\omega}/dH_z=V_{AH}/(-H_k+H_x)$ (see Fig. S1(b)).

The values of $H_k$ we determined for both the PMA (P1-P8) and the IMA (I1-I4) bilayers are plotted in Fig. S1(c). $K_s$ was then determined following the relation $H_k =$ $-4\pi M_s + 2K_v/M_s + 2K_s/M_st$, where $M_s \approx 1330$ emu/cm³ for each sample, and $t$ and $K_v$ are the Co thickness and volume (magneto-elastic and magneto-crystalline) anisotropy.

We measured the anomalous Hall resistivity $\rho_{AH} = V_{AH}t/I_{Co}$, where $I_{Co}$ is the current in the Co layer. The anomalous Hall conductivity is determined as $\sigma_{AH} = V_{AH}L/V_{in}W\rho_{Co}$, where $L$ and $W$ are the length and the width of the Hall bars, respectively.

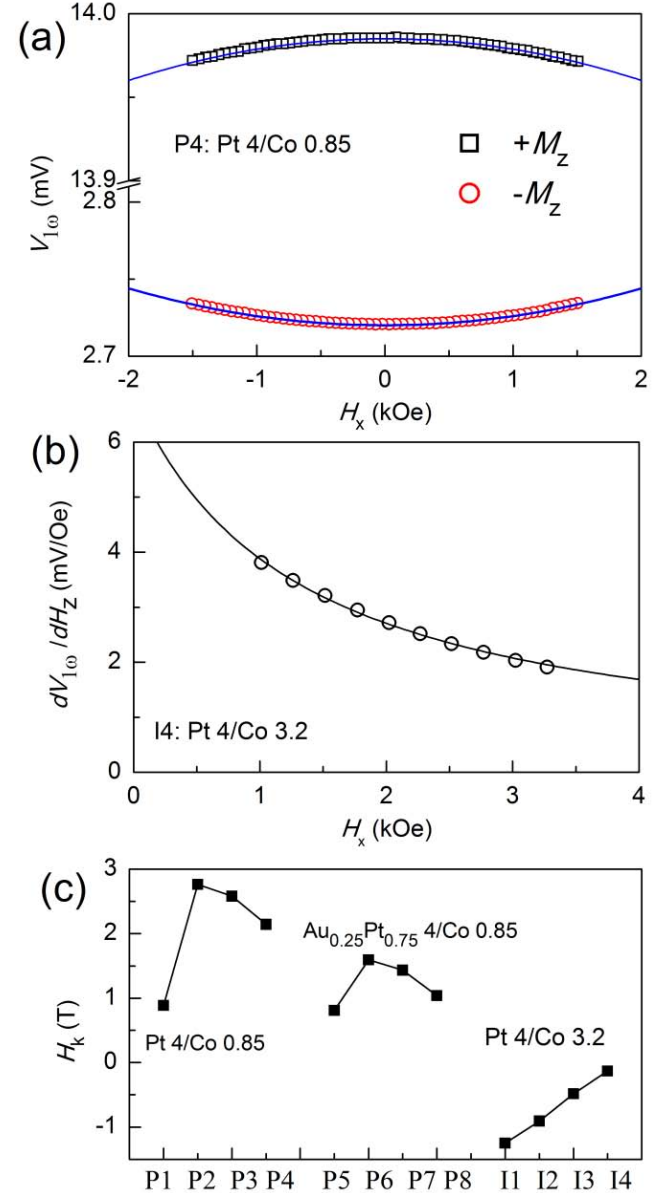

Fig. S1. (a) $H_x$ dependence of $V_{1\omega}$ of the Pt 4/Co 0.85 bilayer annealed at 400 °C for 1 hour (P4) for both $\pm M_z$ cases. The solid lines are best fits to $V_{1\omega}= \pm V_{AH}(1-H^2/2H_k^2)$; (b) $H_x$ dependence of $dV_{1\omega}/dH_z$ for the Pt 4/Co 3.2 bilayer annealed at 450 °C for 1 hour (I4). The solid line is a best fit to $dV_{1\omega}/dH_z=V_{AH}/(-H_k+H_x)$; (c) $H_k$ of the different the magnetic bilayers P1-P8, and I1-I4.

## 3. Annealing effects on Pt/Co and Pt/Hf/Co interfaces

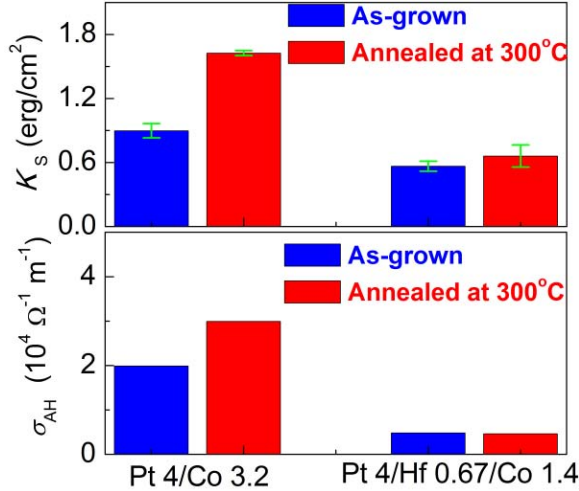

Fig. S2. Comparison of the annealing effects on the interfacial magnetic anisotropy energy density ($K_s$) and the anomalous Hall conductivity ($\sigma_{AH}$) for Pt 4/Co 3.2 and Pt 4/Hf 0.67/Co 1.4. The annealing-induced enhancement of $K_s$ and $\sigma_{AH}$ is significant for Pt/Co interface, while minimal for Pt/Hf/Co interfaces. This is consistent with the spin-orbit proximity effect being the mechanism of interfacial spin-orbit coupling at the HM/Co interfaces.

## 4. Spin-orbit torque efficiencies and resistivities

We determined the SOT efficiencies for the HM/Co bilayers by harmonic response measurements. For the IMA Pt/Co samples, the in-plane angle ($\varphi$) dependence of the out-of-phase second harmonic voltage response $V_{2\omega}$ is given by $V_{2\omega} = V_a\cos\varphi + V_p\cos2\varphi\cos\varphi$, where $V_a = -H_{DL}V_{AH}/2(H_k+H_{in}) + V_{ANE}$, $V_p = -V_{PH}H_{FL}/2H_{in}$, and $V_{PH}$, $V_{ANE}$, and $H_{in}$ are the planar Hall voltage, the anomalous Nernst effect voltage, and the applied in-plane magnetic field, respectively, and $\varphi$ is the angle between $H_{in}$ and the bias current direction (Fig. S3(a)). The $V_a$ and $V_p$ contributions to $V_{2\omega}$ were separated out by fitting the $\varphi$ dependence of $V_{2\omega}$ as measured for different $H_{in}$ (Fig. S3(b)). $H_{DL}$ and $H_{FL}$ were obtained from the slopes of the linear fits of $V_a$ vs $-V_{AH}/2(H_k+H_{in})$ and $V_p$ vs $-V_{PH}/2H_{in}$ (Fig. S3(c)).

For both PMA and IMA samples, the dampinglike and fieldlike SOT efficiencies *per unit bias current density* are $\xi^j_{DL(FL)} = 2e\mu_0M_stH_{DL(FL)}/\hbar j_e$, with $H_{DL(FL)}$, $e$, $\mu_0$, $t$, $\hbar$, and $j_e = V_{in}/L\rho_{HM}$ being the damping-like (field-like) effective spin torque field, the elementary charge, the permeability of vacuum, the ferromagnetic layer thickness, the reduced Planck constant, and the charge current density, respectively. For the PMA HM/Co samples, $H_{DL(FL)} = -2\frac{\partial V_{2\omega}}{\partial H_{x(y)}}/\frac{\partial^2 V_{1\omega}}{\partial^2 H_{x(y)}}$, where $V_{2\omega}$ was measured as a function of in-plane bias fields $H_x$ and $H_y$ (see Figs. S3(d) and S3(e)). The damping-like and field-like SOT efficiencies *per unit applied electric field* are calculated as $\xi^E_{DL(FL)}=\xi^j_{DL(FL)}/\rho_{HM}$.

We note that in analyzing our out-of-plane harmonic-response results we do not employ the so-called "planar Hall correction (PHC)" which was developed to account for magnetoresistance effects (e.g. anisotropic magnetoresistance in the FM and spin Hall magnetoresistance in the HM) in the determination of SOTs in PMA HM/FM bilayers [1]. Although there are only limited discussions in the literature about the "PHC" problem, this is an issue that is rather well known in the spintronics research community [2-4]. Experimentally, whether or not to use the "PHC" is not of any practical concern when the ratio $\zeta = V_{PH}/V_{AH}$, is not larger than ~0.1 as this correction has negligible influence on the value of $H_{DL}$ or $\xi^j_{DL}$ for a PMA sample. However, when $\zeta$ is larger than typically > 0.2, we find that this "correction" can lead to unphysical magnitudes and even sign reversals for the extracted values of the SOT efficiencies, with large discrepancies compared to other measurement methods. To illustrate the invalidity of the "PHC", we show a few examples in Table S1. For $\xi^j_{DL}$, the application of the "PHC" leads to unreasonably large values of $\xi^j_{DL}$ for Pt/Co and $Au_{0.25}Pt_{0.75}/Co$ and an unphysical sign change of $\xi^j_{DL}$ for Pd/Co. The "PHC" also leads to an exceptionally large value of $\xi^j_{FL}$ for Pd/Co and Pt/Co and a sign reversal for Pt/Co. Compared to the values of $\xi^j_{DL}$ obtained from harmonic response measurements on the IMA samples, applying the "PHC" to the PMA data produces large discrepancies. On the other hand, neglecting the PHC for the PMA samples gives $\xi^j_{DL}$ results that are in close accord with the results from the IMA samples with the same FM/HM components and the similar resistivities and thicknesses [4]. (Table S1 shows that the field-like torques in the PMA HM/Co samples can differ from the corresponding IMA samples even without the PHC correction, but we take this as a real physical difference, because the Co layer in the PMA samples is thinner than the spin dephasing length, so that $\xi^j_{FL}$ is generated by spin scattering at both the HM/Co and the Co/MgO interfaces[5], while for the IMA samples the fieldlike term is only sensitive to spin scattering at the HM/Co interface.) While we conclude that applying the "PHC" to PMA samples gives erroneous results, if we do apply this to our data corresponding to different annealing protocols, the anti-correlation of the strength of the damping-like SOT efficiency with ISOC for the PMA samples is not affected, although the amplitude of the damping-like SOTs are unreasonably high in all cases.

The HM resistivity of each sample with the HM thickness $d$ of 4 nm was determined by measuring the conductance enhancement of the corresponding stacks ($d$ =4 nm) with respect to a $d$ = 0 stack. Figure S3(f) shows the evolution of the resistivity of the HM layers ($\rho_{HM}$) and Co layers ($\rho_{Co}$) with enhanced annealing, which we tentatively attribute to atomic structure changes, for instance, strain relaxation. Interestingly, $\rho_{HM}$ is larger in the PMA Pt/Co than in the IMA Pt/Co bilayers, which is mostly likely due to enhanced interfacial scattering. Overall the increase in $\rho_{HM}$ with annealing is relatively small and is unlikely to be indicative of any change of the internal intrinsic spin Hall conductivity of Pt or $Au_{0.25}Pt_{0.75}$ layers. The Fermi surface topology and Berry phase of fcc metals are well-known to be robust to structural impurity and defects. We also note that the change in $\rho_{HM}$ is monotonic with the annealing steps for all the three sample series. Therefore we conclude that the spin diffusion length of the HM layers, which is expected to be determined by the Elliott-Yafet spin-relaxation mechanism [6,7] should also decrease monotonically with annealing. Thus a change in the spin diffusion length cannot readily explain the non-monotonic variation of the spin-orbit torques with annealing.

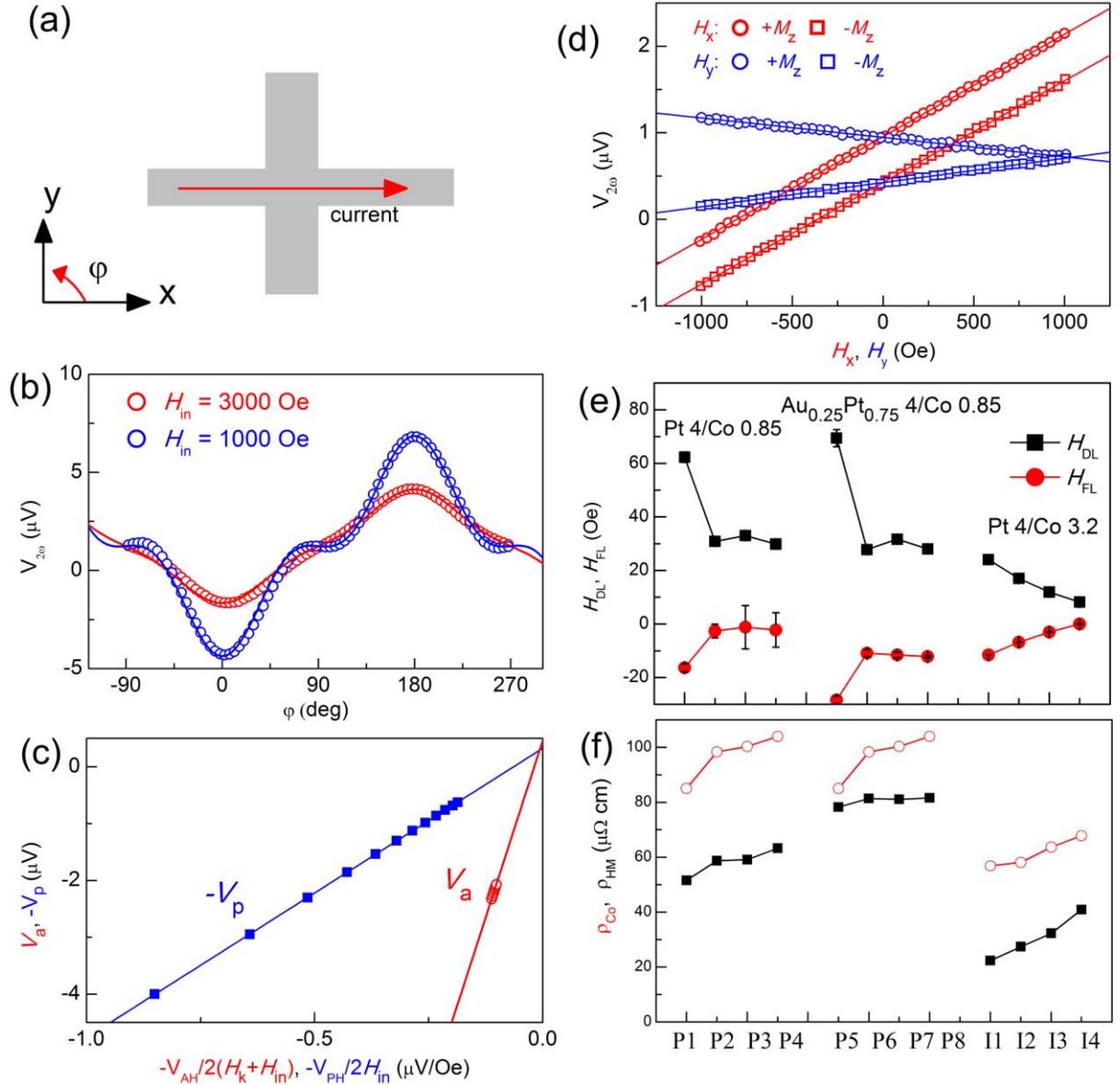

Fig. S3. (a) Schematic showing the measurement coordinate system. (b) $\varphi$ dependence of the second harmonic voltage $V_{2\omega}$ for the Pt 4/Co 3.2 as-grown bilayer (I1) as measured under in-plane bias fields of 1000 and 3000 Oe. The solid lines represent the best fits of the data to relation $V_{2\omega}=V_a\cos\varphi+V_p\cos2\varphi\cos\varphi$. (c) $V_a$ and $V_p$ plotted as a function of $-V_{AH}/2(H_k+H_{in})$ and $-V_{PH}/2H_{in}$ for the Pt 4/Co 3.2 as-grown bilayer (I1). The solid lines are linear fits. (c) The second harmonic voltages for the PMA Pt 4/Co 0.85 as-grown bilayer plotted as a function of in-plane bias fields, $H_x$ (red), and $H_y$ (blue), respectively. (d) The determined $H_{DL}$ and $H_{FL}$. (f) Evolution of the resistivity of the HM layers ($\rho_{HM}$, black squares) and Co layers ($\rho_{Co}$, red circles) with enhanced annealing.

Table S1. Comparison showing the invalidity of the "planar Hall correction" ("PHC") when measuring spin-orbit torques using the PMA harmonic response method. Here, we first compare the harmonic response results for the spin torque efficiency per current density $\xi_{DL(FL)}$ for the PMA samples Pt 4/Co 0.75, Pd 4/Co 0.64, and Au$_{0.25}$Pt$_{0.75}$ 4/0.8 without and with the "PHC". Then we compare to measurements on the IMA samples (Pt 4/Co 1.4, Pd 4/Co 0.94, and Au$_{0.25}$Pt$_{0.75}$ 4/1.4) with the similar HM resistivities. The PMA and IMA results for $\xi_{DL}^{j}$ are in good agreement if the planar Hall correction is not applied and they disagree if it is applied. The PHC for the PMA samples modifies the spin torque efficiency as $\xi_{DL(FL)}^{j}$(With PHC) = $(\xi_{DL(FL)}^{j} + 2\zeta \xi_{DL(FL)}^{j})/(1-4\zeta^2)$, where $\zeta = V_{PH}/V_{AH}$.

|  | $\zeta = V_{PH}/V_{AH}$ | PMA sample No "PHC" | | PMA sample With "PHC" | | IMA sample | |
|---|---|---|---|---|---|---|---|
|  |  | $\xi_{DL}^{j}$ | $\xi_{FL}^{j}$ | $\xi_{DL}^{j}$ | $\xi_{FL}^{j}$ | $\xi_{DL}^{j}$ | $\xi_{FL}^{j}$ |
| Pt/Co | 0.31 | 0.21 | -0.049 | 0.29 | 0.13 | 0.19 | -0.046 |
| Pd/Co | 0.56 | 0.07 | -0.050 | -0.1 | -0.16 | 0.06 | -0.0002 |
| Au$_{0.25}$Pt$_{0.75}$/Co | 0.33 | 0.30 | -0.12 | 0.39 | -0.14 | 0.32 | -0.020 |

## 5. ISOC variations are unrelated to increased interfacial intermixing

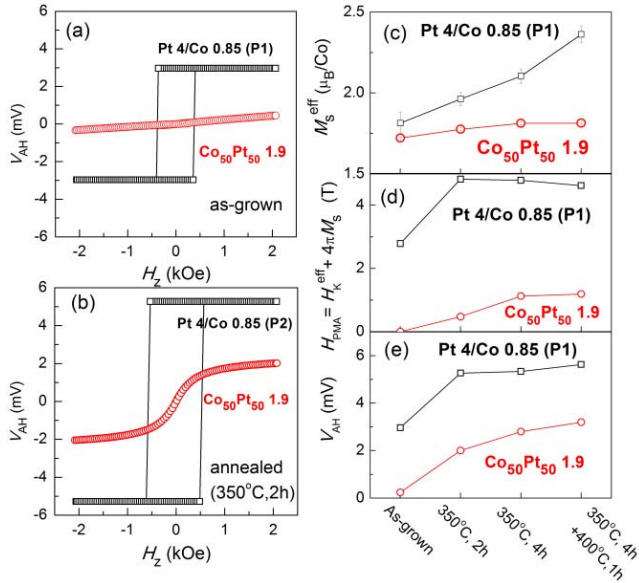

Fig. S4. Comparison of the anomalous Hall voltage ($V_{AH}$) hysteresis loops for (a) the as-grown and (b) the annealed (350 °C, 2h) Pt 4/Co 0.85 bilayer (black squares) and the alloy control sample Co$_{50}$Pt$_{50}$ 1.9 (red circles). (c) Effective saturation magnetization ($M_s^{eff}$) determined by assuming the measured magnetic moments are solely from the Co atoms in the magnetic layer, (d) perpendicular anisotropy field ($H_{PMA}$), and (e) $V_{AH}$ for the Pt 4/Co 0.85 and the Co$_{50}$Pt$_{50}$ 1.9 samples plotted as a function of annealing conditions. For (a), (b), and (e), $V_{AH}$ was measured with a constant electrical field $E \approx 66.7$ kV/m along the Hall bar. All the measurement were performed at 300 K.

In this section, we show that the variations we measure in the perpendicular magnetic anisotropy energy density and the anomalous Hall conductivity cannot be explained by intermixing at the HM/Co interfaces. In Fig. S4, we compare the magnetic behaviors of the Pt 4/Co 0.85 bilayers (P1-P4) with a control sample Si/SiO$_2$/Ta 1/Pt 3.6/Co$_{50}$Pt$_{50}$ 1.9/MgO 2/Ta 1.5 that was sputter deposited with similar growth parameters as those used for other samples in the paper. Here we use the Co$_{50}$Pt$_{50}$ alloy layer as a representative for Co-Pt alloys that could potentially be induced by any intermixing. The thickness of 1.9 nm is to simulate the thickness for the 0.85 nm Co (111) layer being fully alloyed with Pt (lattice constants for $L1_0$-Co$_{50}$Pt$_{50}$ are $a$ = 3.803 Å, $c$ = 3.701 Å; for fcc Co, $a$ = 3.545 Å). As indicated by the sweeps of anomalous Hall voltage ($V_{AH}$) vs applied magnetic field in Fig. S4 (a) and (b), for both the as-grown and 300 °C-annealed cases, the alloy control sample Co$_{50}$Pt$_{50}$ 1.9 shows a non-hysteretic gradual magnetization switching in the film normal direction, a signature for the film normal being the magnetically hard axis. This is in striking contrast to the case of the as-grown and annealed PMA Pt/Co and Au$_{0.25}$Pt$_{0.75}$/Co bilayers (sample sets P1-P8), where the magnetization switching is very sharp, the $V_{AH}$ hysteresis loops are square, and the coercivities are large (410 Oe and 570 Oe for P1 and P2, respectively). We emphasize that such degradation of PMA in real Co-Pt alloy samples compared to the Pt/Co bilayers or multilayers is widely observed and is consistent with previous experimental reports that interfacial intermixing, if substantial, degrades PMA at Pt/Co interface [8-10]. The reason is that chemical disorder in alloys is very difficult to remove, even though the ideal $L1_0$-Co$_{50}$Pt$_{50}$, which is in fact a Pt/Co superlattice stacked along (001), is expected to have strong uniaxial perpendicular magneto-crystalline anisotropy.

Figure S4 (c)-(e) further compares the Pt 4/Co 0.85 bilayer and the alloy control sample Co$_{50}$Pt$_{50}$ 1.9 in terms of the magnitudes of the effective saturation magnetization ($M_s^{eff}$), the perpendicular anisotropy field ($H_{PMA}$), and the anomalous Hall voltage ($V_{AH}$) for different annealing conditions. Here $M_s^{eff}$ is determined by assuming the measured magnetic moments are solely from the Co atoms in the magnetic layer; $H_{PMA}$ is the sum of the effective perpendicular anisotropy field ($H_k^{eff}$) and demagnetization field ($4\pi M_s^{eff}$); and $V_{AH}$ for each sample is measured with an alternating electric field with a magnitude $E = V_{in}/L = 66.7$ kV/m applied along the 5×60 um$^2$ Hall bars. For each annealing condition we used for this work, the alloy sample has much smaller values of $M_s^{eff}$, $H_{PMA}$ and $V_{AH}$ than the Pt 4/Co 0.85 bilayer. The small magnitude and weak annealing dependence of $M_s^{\text{eff}}$ for the alloy sample (i.e.1.72-1.8 $\mu_B$/Co) obviously fail to explain the initially large and then quickly increasing $M_s^{\text{eff}}$ we observed for all the bilayer samples in this work (i.e. 1.81-2.36 $\mu_B$/Co). As discussed in detail in [11], at room temperature an occurrence of any other degree of Co-Pt intermixing should also lead to either a dead layer (paramagnetic A1, $L1_2$–CoPt$_3$) or a reduction of effective magnetization of Co ($L1_2$–Co$_3$Pt, ~1.72 $\mu_B$/Co). We therefore conclude that any significant Pt-Co intermixing at the Pt/Co interface should degrade the effective magnetization and thus the magnetic proximity effect and can be excluded in our Pt/Co (111) or Au$_{0.25}$Pt$_{0.75}$/Co (111) bilayers. Similarly, $H_{\text{PMA}}$ for the alloy sample Co$_{50}$Pt$_{50}$ 1.9 (Fig. S4(d)) is at least ~5 times smaller than that of the Pt 4/Co 0.85 bilayers (i.e. 2.76-5.6 T). The measured $V_{\text{AH}}$ for the alloy sample Co$_{50}$Pt$_{50}$ 1.9 (Fig. S4(e)) is a factor of ~20 smaller than that of Pt/Co layers before annealing and about a factor of 2 smaller afterwards,, and is therefore too small to account for the large $V_{\text{AH}}$ values in Pt/Co (111) bilayers ($V_{\text{AH}}$ is 2.97, 5.27, 5.34, and 5.63 mV for P1-P4, respectively). Note that such a large $V_{\text{AH}}$ difference between the Co-Pt alloy and Pt/Co (111) bilayer cannot be attributed to the different ferromagnetic layer thickness. In this measurement geometry, the value of $V_{\text{AH}}$ is not influenced by the evaluation of the ferromagnetic layer thickness because it is given as $V_{\text{AH}} = Ew\rho_{\text{AH}}/\rho_{xx}$ with $w$=5 um, $\rho_{\text{AH}}$, $\rho_{xx}$, and $\rho_{\text{AH}}/\rho_{xx}$ being the Hall bar width, the anomalous Hall resistivity, the longitudinal resistivity, and the anomalous Hall angle of the ferromagnetic layer. The results for $M_s^{\text{eff}}$, $H_{\text{PMA}}$ and $V_{\text{AH}}$ for the bilayers in the main text are therefore all consistent with the occurrence of a strong magnetic proximity effect, but are clearly contrary to the results for Co-Pt alloying [11].

In particular, the PMA and $V_{\text{AH}}$ for all the samples, i.e. PMA Pt/Co and Au$_{0.25}$Pt$_{0.75}$/Co, and IMA Pt/Co, *increase* substantially right after the first annealing. This excludes the possibility of intermixing occurrence after the first annealing step because Pt-Co intermixing *decreases* PMA and $V_{\text{AH}}$ as we show in Fig. S4. Instead, the increase of PMA and $V_{\text{AH}}$ indicate that the interface is overall more ordered as the result, since chemical disorder is expected to degrade the PMA [9,10]. The *non-monotonic* changes of the interfacial magnetic anisotropy energy density and $V_{\text{AH}}$ for the PMA bilayers of Pt/Co and Au$_{0.25}$Pt$_{0.75}$/Co strongly disagree with any mechanisms based on annealing-induced intermixing as the intermixing is unlikely to vary with annealing in a non-monotonic manner.

## 6. SOT variations are unrelated to increased interfacial intermixing

The annealing-induced variations of the SOT efficiencies of our samples also cannot be explained by increased interfacial intermixing. To demonstrate this point directly we produced two additional Pt/Co samples, one with an insertion layer of Pt$_3$Co 0.2 nm/Co$_{50}$Pt$_{50}$ 0.2 nm/Co$_3$Pt 0.2 nm at the Pt/Co interface as a simulation of a possible intermixing upon annealing and one without this insertion layer. As shown in Fig. S5, the insertion of the alloy layer at the interface has no discernable effect on $H_{\text{DL}}$ compared to the sample without the insertion. The almost identical $H_{\text{DL}}$ values for the Pt 4/Co 2.3 samples with and without the alloy insertion layer indicate that even if there were interfacial intermixing in our annealed samples this would not cause a large change in $\xi_{\text{DL}}^{\text{E}}$. When the thin Pt-Co alloy insertion layer is present, the ISOC as indicated by $H_K$ is reduced, again indicating that a strongly intermixed Pt-Co interface does not, of itself, result in a stronger ISOC. $V_{\text{AH}}$ is also slightly reduced in the sample with the thin Pt-Co-insertion layer, consistent with a decrease of the ISOC and with the Pt-Co alloys having negligibly small $V_{\text{AH}}$ contribution.

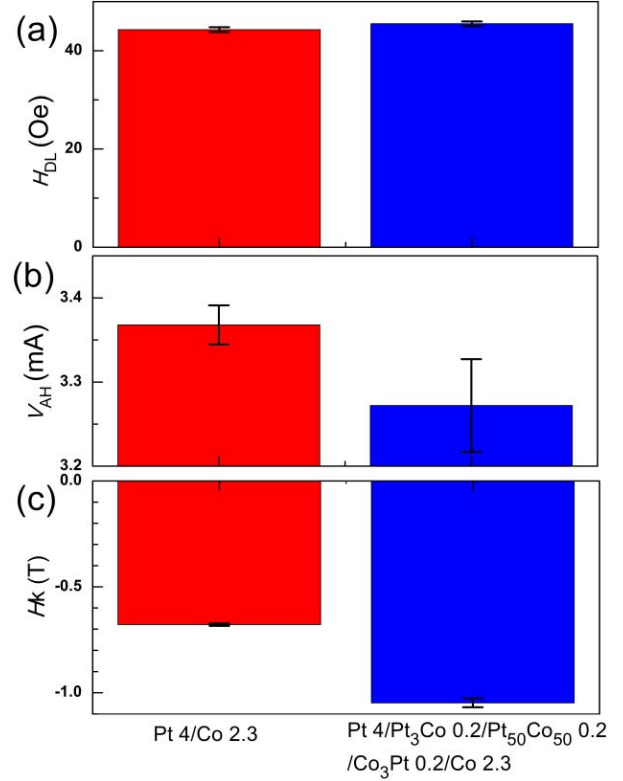

Fig. S5 (a) Dampinglike spin-torque fields ($H_{\text{DL}}$), (b) anomalous Hall voltages ($V_{\text{AH}}$), and (c) effective anisotropy fields for Pt 4/Co 2.3 samples without (red) and with (blue) alloy spacer Pt$_3$Co 0.2/Co$_{50}$Pt$_{50}$ 0.2/Co$_3$Pt 0.2. In (a) and (b), the electric bias field ≈66.7 kV/m.

With respect to the dampinglike SOT efficiency, even if annealing did generate interfacial intermixing we would expect this to *enhance* $\sigma_{\text{SH}}$ and the dampinglike spin-torque efficiency per electric field ($\xi_{\text{DL}}^{\text{E}}$) rather than to produce the decreases reported in the main text. Diffusion of Co into Pt should increase the electron scattering rate and the Pt resistivity $\rho_{xx}$ (we determine the resistivity of the Co$_{50}$Pt$_{50}$ sample to be ~90 μΩ cm, higher than that for Pt in Fig. S3(f)) and thus reduce the spin diffusion length $\lambda_s$ of the Pt layer due to the Elliot–Yafet spin relaxation mechanism ($\lambda_s \propto 1/\rho_{xx}$). Thus a direct consequence of substantial intermixing would be an *increase* of $\xi_{\text{DL}}^{\text{E}}$ because $\xi_{\text{DL}}^{\text{E}}(d) = \xi_{\text{DL}}^{\text{E}}(\infty)(1-\sec h(d/\lambda_s))$ with the Pt thickness $d = 4$ nm. Doping magnetic impurities into Pt has also been found to introduce strong additive skew scattering, which can increase the total spin Hall conductivity and $\xi_{DL}^{E}$ [12]. As shown in the main text, for the in-plane Pt/Co sample, $\xi_{DL}^{E}$ decreases monotonically with annealing, which is sharply contrary to an occurrence of a significant intermixing at the HM/Co interface.

## 7. Direct structural characterization of the HM/Co interfaces

Even though we have already shown that neither the ISOC variations nor the SOT variations are due to annealing-induced interfacial intermixing, it is still interesting to directly characterize whether or not there is intermixing at the HM/Co interfaces after annealing. We first performed x-ray reflectivity measurements. In the incident-angle-dependent x-ray reflectivity measurements, the oscillation strength and attenuation length reflect the interface sharpness. As can be seen from Figs. S6(a)-(c), the oscillation strength and attenuation length both indicate good interface sharpness in the as-grown state, and the quality of these signals is improved after the final-step annealing. This indicates that annealing-induced intermixing at the interfaces of the bilayer samples should be minimal.

We also carried out chemical depth profile measurements with secondary ion mass spectrometry (SIMS) to independently characterize the HM/Co interfaces. In order to minimize the etching-induced diffusion and to maximize the depth resolution, we used low-energy oblique argon-beam etching (accelerating voltage = 150 V, 20° degrees away from the sample surface). As shown in Fig. S7(a), if we compare the etching time when the Co and Pt signals reach their maximums, we find that the Co and Pt signals reach their maximums at the same time for the reference alloy sample $Co_{50}Pt_{50}$, whereas the peaks in the Co and Pt signals are well separated for all the Pt/Co bilayers (see Figs. S7(b)-(e)). In both the IMA and the PMA cases, the etching time for the maximization and disappearance of Pt signal for the as-grown Pt/Co bilayers are the same as for the annealed ones, indicating that there is at most minimal Pt diffusion into Co. For both IMA and PMA samples, the Co signal for the as-grown and annealed samples reaches the maximum after the same etching time, suggesting no annealing-enhanced intermixing between the MgO and the Co layers (see Fig. S8 (a)-(d)). However, we note that the Co tails of the two annealed samples (P4 and I4) decay slightly more slowly than those of the corresponding as-grown bilayers (P1 and I1) despite their same disappearance time (see Figs. S8(e) and (f)). This could indicate a small amount of Co diffusion into the top layer of Pt. As discussed above, any inter-diffusion of Co must be limited in amount and extent to be consistent with the experimental results for the interfacial spin-orbit coupling (ISOC) strength (PMA, $V_{AH}$), $M_s^{eff}$, and the oscillation strength and decay length of x-ray reflectivity curves, because all will be substantially degraded if there was significant intermixing of the Co and Pt layers. Therefore we conclude that although there may be some diffusion of Co atoms into Pt layer, such diffusion has a negligible affect on the ISOC and spin-torque properties.

Finally we note that in literature there is disagreement on the thermal stability of the Pt/Co interfaces. MeGee *et al*. find sputtered amorphous Pt/Co samples are more stable than those epitaxially grown on single-crystal substrate by molecular-beam epitaxy [10], which suggests that the thermal stability can be sensitive to interface details. K. Yakushiji *et al*. [13] find Pt/Co superlattices are thermally stable to at least above 370°C. In contrast to the assumption that annealing can enhance intermixing at Pt/Co interfaces, some experiments reveal that high temperature annealing *separates* Co out from a disordered Co-Pt mixture and improve the magnetic anisotropy of Co/Pt superlattices [9].

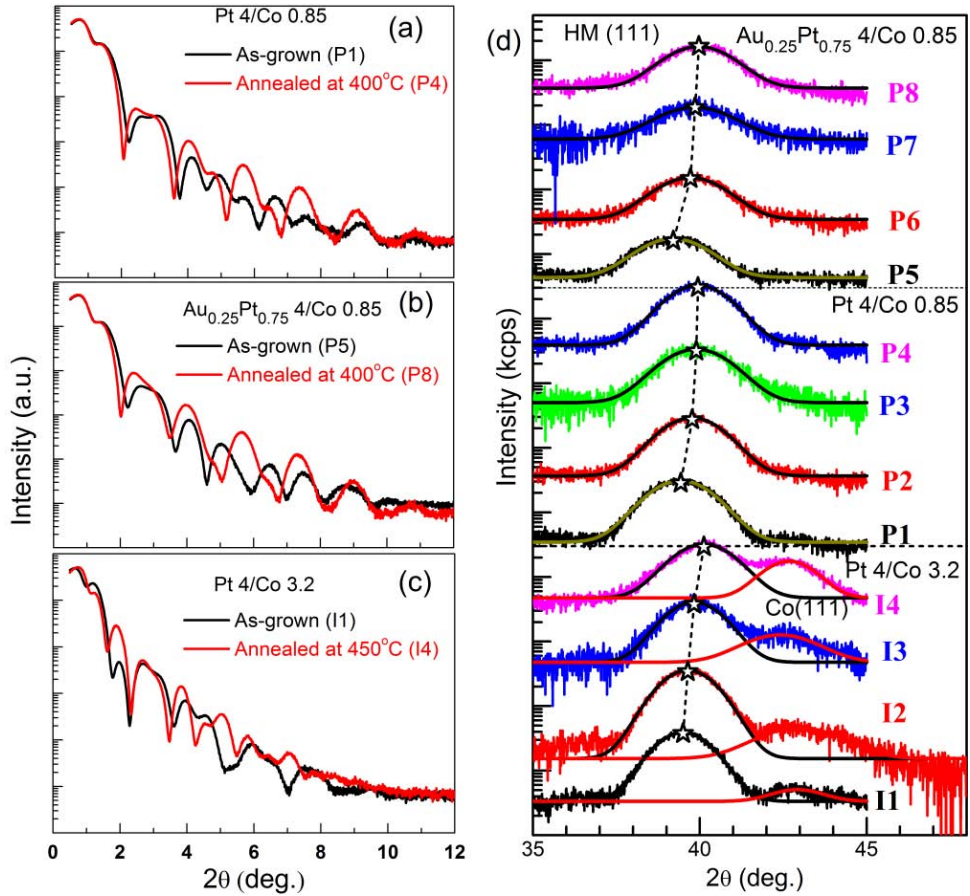

Fig. S6 X-ray reflectivity curves for (a) as-grown (P1) and annealed (P4) Pt 4/Co 0.85; (b) as-grown (P5) and annealed (P8) $Au_{0.25}Pt_{0.75}$ 4/Co 0.85; (c) as-grown (I1) and annealed (I4) Pt 4/Co 3.2. (d) X-ray diffraction $\theta$-$2\theta$ patterns for the Pt 4/Co 0.85 (P1-P4), the $Au_{0.25}Pt_{0.75}$ 4/Co 0.85 (P5-P8), the Pt 4/Co 3.2 bilayer samples (I1-I4) with different annealing steps. The solid lines refer to best two-peak fits of the measured curves. As indicated by the stars in (d), the shift of the HM (111) peak with enhanced annealing indicates the relaxation of the strains in the bilayers.

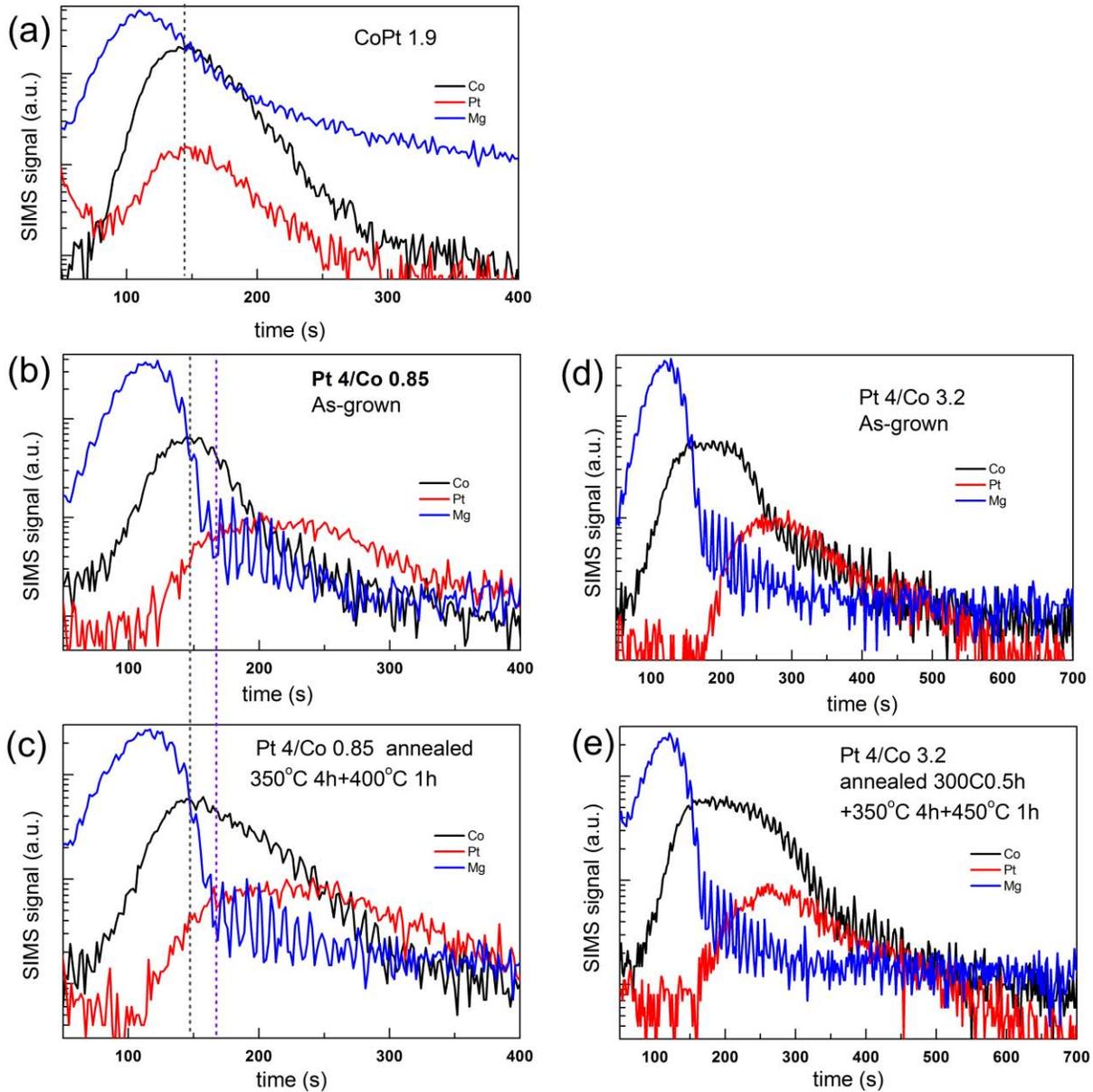

Fig. S7 SIMS measurements of the depth profile of Co, Pt, and Mg elements for (a) the alloy control sample $Co_{50}Pt_{50}$ 1.9 (corresponding to fully alloyed Pt 0.85/Co 0.85), (b) as-grown Pt 4/Co 0.85 (corresponding to P1), (c) Pt 4/Co 0.85 annealed at 350 °C 4 h and 400 °C 1 h (corresponding to P4), (d) as-grown Pt 4/Co 3.2 (corresponding to I1) and (e) Pt 4/Co 3.2 annealed at 300 °C 0.5 h, 350 °C 4 h, and 400 °C 1 h (corresponding to I4). All the samples are seeded with a 1 nm Ta and protected with MgO 2/Ta 1.5. Here we use low-energy oblique argon-beam etching (accelerating voltage of 150 V, 20° degrees away from the sample surface) to minimize the etching-induced heating and diffusion and to maximize the depth resolution. If we compare the etching time when the Co and Pt signals reach their maximums, we find that the Co and Pt peaks overlaps very well for the alloy control sample $Co_{50}Pt_{50}$, whereas they are well separated for all the Pt/Co bilayers. As an example, we use the two dashed lines for (a)-(c) to indicate the etching time when the Co and Pt signals reach maximums.

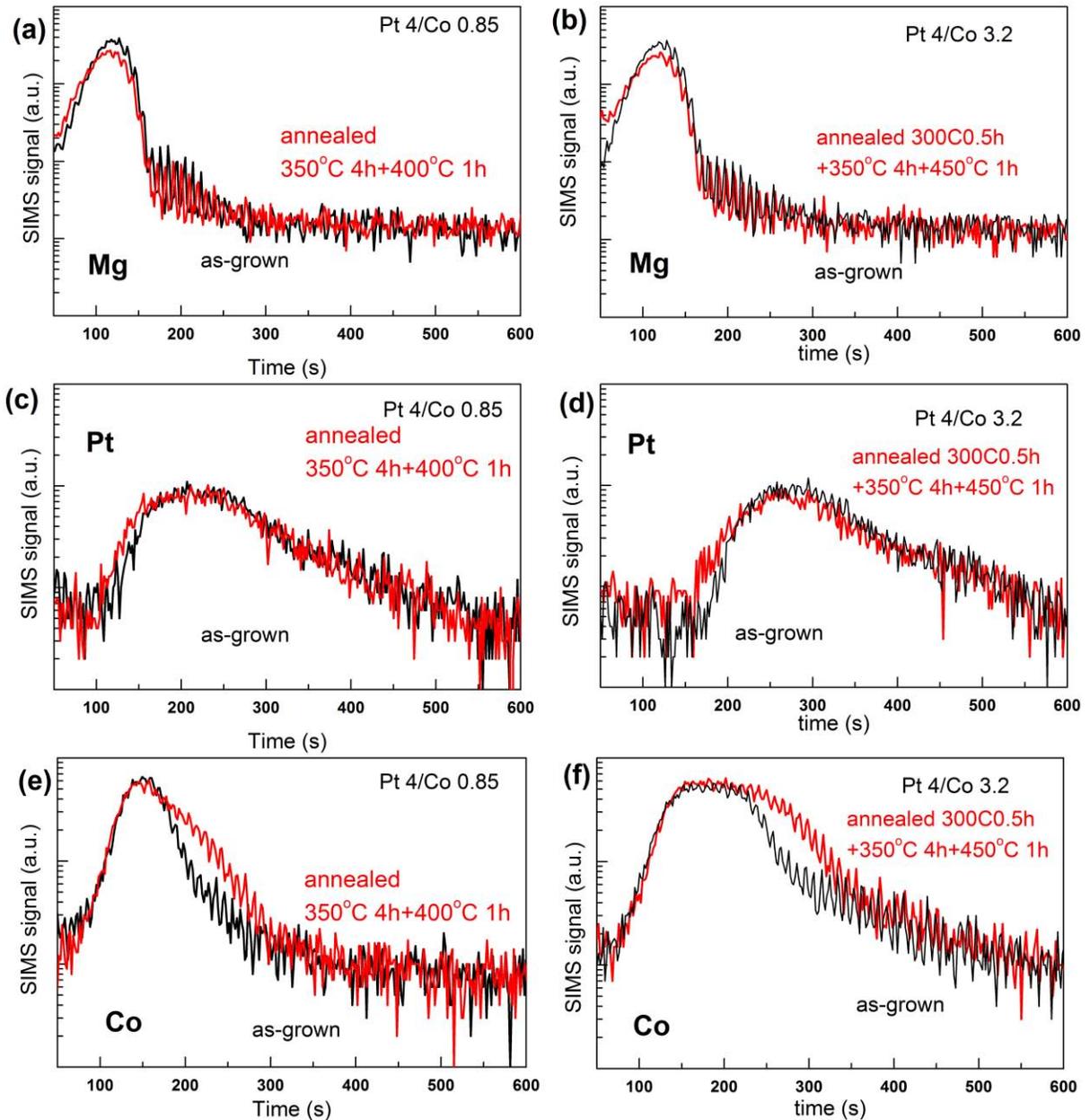

Fig. S8 Comparison of the SIMS signals of Mg, Pt, and Co elements for as-grown (black) and annealed (red) Pt/Co bilayers as a function of the etching time. (a),(b) Mg; (c),(d) Pt; (e),(f) Co. Within the resolution and stability of the instrument, Mg and Pt don't show evident diffusion; while the slower decays of the Co signals for the annealed samples are likely suggestive of a small amount of diffusion of Co into Pt.

## 8. Strain variation

Figure 6(d) shows the variation of the x-ray diffraction $\theta$-$2\theta$ patterns for the Pt 4/Co 0.85 (P1-P4), the $Au_{0.25}Pt_{0.75}$ 4/Co 0.85 (P5-P8), the Pt 4/Co 3.2 bilayer samples (I1-I4) as a function of annealing steps. The shift of the HM (111) peak with annealing indicates relaxation of the strains in the bilayers. For the PMA bilayers, the HM/FM peaks shifts to higher angle after the first annealing step, and then remains relatively constant with further annealing. For the IMA samples where the Co is 3.2 nm thick, the Pt peak shifts monotonically to higher angles, while the Co peak does the opposite. The different strain variation of the PMA and IMA bilayers is likely related to the different variations of interfacial ISOC strength that the two different types of bilayers exhibit (see Fig. 1 in the main text).

---

**References**

[1] M. Hayashi, J. Kim, M. Yamanouchi, and H. Ohno, Quantitative characterization of the spin-orbit torque using harmonic Hall voltage measurements, Phys. Rev. B 89, 144425 (2014).
[2] Y.-C. Lau and M. Hayashi, Spin torque efficiency of Ta, W, and Pt in metallic bilayers evaluated by harmonic Hall and spin Hall magnetoresistance measurements, Jpn. J. Appl. Phys. 56, 0802B5 (2017).